\documentclass[aps,prl,twocolumn,scriptaddress,floatfix]{revtex4}
\usepackage{amsmath,amssymb,amsfonts,amsthm,url,xspace,graphicx,psfrag,subfigure,multirow}
\usepackage[pdftitle={The dimension of loop-erased random walk in 3D},
            pdfauthor={David B. Wilson},
            bookmarks=true,bookmarksopen=true,bookmarksopenlevel=2]{hyperref}
\usepackage[all]{hypcap}
\usepackage{cancel}

\newcommand{\sref}[1]{\hyperref[#1]{\S~\ref*{#1}}}
\newcommand{\aref}[1]{\hyperref[#1]{Appendix~\ref*{#1}}}
\newcommand{\lref}[1]{\hyperref[#1]{Lemma~\ref*{#1}}}
\newcommand{\tref}[1]{\hyperref[#1]{Theorem~\ref*{#1}}}
\newcommand{\cref}[1]{\hyperref[#1]{Corollary~\ref*{#1}}}
\newcommand{\fref}[1]{\hyperref[#1]{Figure~\ref*{#1}}}
\newcommand{\pref}[1]{\hyperref[#1]{Proposition~\ref*{#1}}}

\def\clap#1{\hbox to 0pt{\hss#1\hss}}

 \newcommand{\MRhref}[2]{\href{http://www.ams.org/mathscinet-getitem?mr=#1}{MR#2}}

\makeatletter
\def\@strippedMR{}
\def\@scanforMR#1#2#3\endscan{%
  \ifx#1M\ifx#2R\def\@strippedMR{#3}%
  \else\def\@strippedMR{#1#2#3}%
  \fi\fi}
\def\@rst #1 #2other{#1}
\newcommand\MR[1]{\relax\ifhmode\unskip\spacefactor3000 \space\fi
  \@scanforMR#1\endscan
  \MRhref{\expandafter\@rst \@strippedMR other}{\@strippedMR}}
\newcommand\MRs[1]{\relax\ifhmode\unskip\spacefactor3000 \space\fi
  \@scanforMR#1\endscan
  \MRhref{\@strippedMR}{\@strippedMR}}
\makeatother

\numberwithin{theorem}{section}

\theoremstyle{definition}
\theoremstyle{definition}
\newcommand{\Z}{\mathbb{Z}}

\newcommand{\eps}{\varepsilon}
\renewcommand{\th}{\ensuremath{^{\text{th}}}\xspace}

\def\rcs $#1: #2 ${\expandafter\def\csname rcs#1\endcsname {#2}}
\rcs $Date: 2008/12/13 00:07:24 $

\begin{document}
\title{The dimension of loop-erased random walk in 3D}
\author{\href{http://dbwilson.com}{David B. Wilson}}
\affiliation{Microsoft Research, Redmond, WA 98052, USA}
\date{August 6, 2010; revised October 7, 2010}

\begin{abstract}
  We measure the fractal dimension of loop-erased random walk (LERW)
  in 3 dimensions, and estimate that it is $1.62400\pm 0.00005$.  LERW
  is closely related to the uniform spanning tree and the abelian
  sandpile model.  We simulated LERW on both the cubic and
  face-centered cubic lattices; the corrections to scaling are
  slightly smaller for the face-centered cubic lattice.
\end{abstract}

\maketitle

\section{Introduction}

Loop-erased random walk (LERW) is the path obtained from a random walk
by erasing loops as they are formed, and was introduced by Lawler.
The paths connecting points in a uniformly random spanning tree are
distributed as loop-erased random walk
\cite{pemantle,majumdar,MR1427525}.  Uniform spanning trees in turn
are closely related to the abelian sandpile model of self-organized
criticality \cite{majumdar-dhar}, and properties of loop-erased random
walk manifest themselves in the avalanches of these sandpiles.  In
particular, the fractal dimension $z$ of LERW is related to the
scaling behavior of topplings in the sandpile models
\cite{priezzhev,KLGP}.

In two dimensions, LERW has dimension $5/4$
\cite{majumdar,kenyon,masson}, and its scaling limit
(with suitable boundary conditions) is known to be SLE$_2$
\cite{LSW}.  In dimensions 4 and higher, LERW has dimension 2, though
with a cube-root log correction in 4D \cite{lawler}.  In dimension 5
and higher, the scaling limit of LERW is Brownian motion.  By
contrast, relatively little is known about LERW in 3D.

Kozma proved that 3D LERW (in certain domains) has a scaling limit
that is invariant under rotations and dilations \cite{kozma}.  The
Hausdorff dimension of 3D LERW is not rigorously known to be
well-defined, but $1< z \leq 5/3$ \cite{lawler-99}.  There have been a
number of estimates of this dimension $z$, sometimes expressed in
terms of $\nu = 1/z$ or $2\nu$.  Guttmann and Bursill estimated
$2\nu=1.600\pm0.006$ in $d=2$ and $2\nu=1.232\pm0.008$ in $d=3$
\cite{guttmann-bursill}.  Bradley and Windwer estimated
$2\nu=1.571\pm0.006$ in $d=2$ and $2\nu=1.230\pm0.003$ in $d=3$
\cite{bradley-windwer}.  Anton predicted $\nu=8/13$ exactly
\cite{anton}.  Agrawal and Dhar estimated $z(d=3)=1.6183 \pm 0.0004$
\cite{agrawal-dhar}.  Fedorenko, Le Doussal, and Wiese gave an
expansion for spatial dimension $4-\eps$:
$$
z(d=4-\eps) = 2 - \frac{\eps}{3} - \frac{\eps^2}{9} + O(\eps^3).
$$
Evaluating this series at $\eps=1$, they estimated $z(d=3)= 1.614 \pm
0.011$ \cite{FDW}.  More recently, Grassberger estimated
$z(d=3)=1.6236\pm0.0004$ \cite{grassberger}, contradicting
Agrawal and Dhar's earlier estimate.  Our estimate is even more
precise: $1.62400\pm0.00005$.  These estimates are summarized
in Table~\ref{tbl:estimates}.

\begin{table}
\begin{center}
\begin{ruledtabular}
\begin{tabular}{cl}
dimension of 3D LERW & reference \\
\hline
 $1<z\leq 5/3$ (rigorous) & Lawler \cite{lawler-99}\\
 $1.623\pm 0.01$& Guttmann \& Bursill \cite{guttmann-bursill} \\
 $1.626\pm 0.004$& Bradley \& Windwer \cite{bradley-windwer} \\
 $13/8$ (conjectured exact) & Anton \cite{anton} \\
 $1.6183 \pm 0.0004$ & Agrawal \& Dhar \cite{agrawal-dhar} \\
 $1.614 \pm 0.011$ & Fedorenko, Le Doussal, \& Wiese \cite{FDW} \\
 $1.6236 \pm 0.0004$ & Grassberger \cite{grassberger} \\
 $1.62400 \pm 0.00005$ & present work \\
\end{tabular}
\end{ruledtabular}
\end{center}
\caption{Estimates of the dimension $z$ of loop-erased random walk in 3 dimensions.}
\label{tbl:estimates}
\end{table}

\section{Simulation design}

Most earlier simulations (with the exception of Agrawal and Dhar's
simulations \cite{agrawal-dhar}) look at the length of the loop erasure
of a random walk run for a large number of time steps.  The starting
point of a loop-erased random walk has different statistical
properties than typical points on the LERW path.  For example, the
winding angle variance at the starting point of LERW is different
than at a typical point \cite{WW,duplantier}.  To measure the
dimension of LERW, we would like to measure the length
of an LERW path without the atypical starting point.

Agrawal and Dhar's simulations instead created loop-erased random
loops, and measured their lengths.  Dhar and Dhar had argued that
adding an edge to a tree creates a loop of size $\ell$ with
probability $\approx \ell^{-2/z}$, and that the next step of an LERW
produces a loop of size $\ell$ with probability $\approx
\ell^{-1-2/z}$ \cite{dhar-dhar}.  Agrawal and Dhar estimated the LERW
dimension $z$ by looking at the sizes of these erased loops.  In their
estimate, it was necessary to ignore both small loops (because of
lattice effects) and large loops (which were influenced by the LERW
stopping condition), effectively leaving fewer length scales with
which to estimate the dimension.

In our simulations, we ran a random walk on an $L\times L \times L$
torus while erasing contractible loops, until a noncontractible loop-erased random loop
was formed, and reported its length.  This random variable is equidistributed to
the following: Generate a uniformly random directed subgraph of the
torus, where each vertex has out-degree 1, conditioned on there being
no contractible cycles.  Every vertex leads to a cycle that winds
around the torus.  Pick a random vertex, find the cycle that it leads
into, and report its length.  By taking many such measurements
for different $L$'s, we estimated the dimension of LERW.

The geometry of the torus affects the length distribution of the
noncontractible loop, but for different $L$'s the effect is the same.
While Agrawal and Dhar had to ignore both small loops and large loops,
we only need to ignore the small $L$'s (because of lattice effects),
so it is easier to see the asymptotic behavior.

\section{Choice of lattice}

Kozma proved that LERW on any 3D lattice converges to the same scaling
limit \cite{kozma}, which is invariant under dilations and rotations,
provided that random walk on the lattice converges to Brownian motion.
To measure the dimension or other properties of LERW, in addition to
using the standard cubic lattice, we also tried the face-centered
cubic (FCC) lattice.  The FCC lattice arises from the densest packing
of spheres in 3D.  Each site is adjacent to 12 nearest neighbors, as
opposed to 6 nearest neighbors in the cubic lattice.  In this sense
the FCC lattice is closer to being isotropic than the cubic lattice,
and since the scaling limit of LERW is isotropic, we might expect that
LERW on the FCC lattice behaves more like the isotropic scaling limit
for smaller values of $L$ than LERW on the cubic lattice.

The most convenient way to simulate on the face-centered cubic lattice
is to use the same $L\times L\times L$ grid as for the cubic lattice,
but with extra edges, where $x$ and $y$ are connected if $x-y$ is one
of $(\pm1,0,0)$, $(0,\pm1,0)$, $(0,0,\pm1)$, $(\pm1,\mp1,0)$,
$(0,\pm1,\mp1)$, or $(\pm1,0,\mp1)$.  In addition to changing the lattice,
this also changes the geometry of the torus, making it a skew torus.
But the geometry of this skew torus is plausibly better than the geometry
of the ordinary torus, since the girth divided by volume$^{1/3}$ is larger.

Our simulations suggest that the face-centered cubic lattice gives
slightly better results than the ordinary cubic lattice for systems
with the same side length $L$.  It would be interesting to see what
effect the choice of lattice has on simulations in higher dimensions,
where there are lattices that are much better (by some measures) than
the hypercubic lattice.

\section{Loop homology}

In order for the measured loop length to be related to the dimension
of LERW, we would need to know that the loops do not wind around the
torus too many times, and in particular that the number of windings
does not grow with the side length $L$ (as it would in $5$ and higher
dimensions).  Therefore we measured the homology of the
noncontractible loop in addition to its length, and found that in fact
it does not grow with $L$ (see Table~\ref{tbl:homology}).

\begin{table}
\begin{ruledtabular}
\scriptsize 
\begin{tabular}{ccccc}
& \normalsize homology & \normalsize $L=64$ & \normalsize $L=1024$ & \normalsize $L=16384$ \\
\hline
\multirow{11}{*}{\normalsize \begin{rotatebox}{90}{cubic lattice}\end{rotatebox} \begin{rotatebox}{90}{cubic torus}\end{rotatebox}}
& 1,0,0 & 0.6242 & 0.6214 & 0.6212 \\
& 1,1,0 & 0.2804 & 0.2813 & 0.2814 \\
& 1,1,1 & 0.0613 & 0.0618 & 0.0619 \\
& 2,1,0 & 0.0163 & 0.0169 & 0.0170 \\
& 2,1,1 & 0.0084 & 0.0087 & 0.0087 \\
& 2,0,0 & 0.0077 & 0.0080 & 0.0080 \\
& 2,2,1 & 0.0007 & 0.0007 & 0.0007 \\
& 2,2,0 & 0.0005 & 0.0006 & 0.0006 \\
& 3,1,0 & 0.0002 & 0.0002 & 0.0002 \\
& 3,1,1 & 0.0001 & 0.0001 & 0.0001 \\
& 3,0,0 & 0.0001 & 0.0001 & 0.0001 \\
\hline
\hline
\multirow{9}{*}{\normalsize \begin{rotatebox}{90}{FCC lattice}\end{rotatebox} \begin{rotatebox}{90}{ skew torus}\end{rotatebox}}
& $1,0,0 $ & 0.8395 & 0.8377 & 0.8375 \\
& $1,1,-1$ & 0.0781 & 0.0784 & 0.0785 \\
& $1,1,0 $ & 0.0674 & 0.0683 & 0.0684 \\
& $2,0,0 $ & 0.0067 & 0.0070 & 0.0070 \\
& $2,1,-1$ & 0.0061 & 0.0063 & 0.0063 \\
& $2,1,0 $ & 0.0012 & 0.0013 & 0.0013 \\
& $1,1,1 $ & 0.0007 & 0.0007 & 0.0008 \\
& $2,2,-1$ & 0.0001 & 0.0001 & 0.0001 \\
& $2,2,-2$ & 0.0001 & 0.0001 & 0.0001 \\
\end{tabular}
\end{ruledtabular}
\caption{Empirical probability distribution of the homology (up to symmetry) of the noncontractible loop.}
\label{tbl:homology}
\end{table}

The fact that the loops with high
probability do not wind around the torus more than $O(1)$ times can be
deduced as follows.  A classical result says that two independent random
walks started distance $R$ apart and run $R^2$ steps in $\Z^3$ have a
constant chance of intersecting.  Lyons, Peres, and Schramm \cite{LPS}
showed that whenever two independent random walks intersect with
constant probability, the second random walk intersects the
loop-erasure of the first with constant probability.  Every time the
random walk winds twice more around the $L\times L\times L$ torus,
there is a constant chance that the second time around it intersects
the loop-erasure of the first time around.  So the number of windings
is stochastically dominated by a geometric random variable, uniformly in~$L$.

\section{Notes on the simulations}

\begin{table*}[t]
\newcommand{\zp}[2]{\tiny$\!\begin{matrix}\ \\[-4pt]\begin{matrix}#1\end{matrix}\\#2\\[-4pt]\ \end{matrix}\!$}
\newcommand{\zpc}[2]{\tiny$\!\begin{matrix}\ \\[-4pt]\cancel{\begin{matrix}#1\end{matrix}}\\#2\\[-4pt]\ \end{matrix}\!$}
\begin{ruledtabular}
\begin{tabular}{ccccccc}
lattice
& $2^{8},\ldots,2^{14}$ 
& $2^{9},\ldots,2^{14}$ 
& $2^{10},\ldots,2^{14}$ 
& $2^{11},\ldots,2^{14}$ 
& $2^{12},\ldots,2^{14}$ 
& $2^{13},2^{14}$ 
\\ \hline
cubic 
&\zpc{1.62393\pm.00001\\0.11712\pm.00006\\}{p=3\times 10^{-22}}
&\zpc{1.62396\pm.00001\\0.11687\pm.00008\\}{p=0.03}
&\zp{1.62397\pm.00001\\0.11677\pm.00011\\}{p=0.4}
&\zp{1.62398\pm.00002\\0.11668\pm.00015\\}{p=0.75}
&\zp{1.62398\pm.00003\\0.1167\pm.0003\\}{p=0.57}
&\zp{1.62399\pm.00006\\0.1166\pm.0005\\}{\text{$p$ undefined}}
\\[9pt]
FCC 
&\zpc{1.62394\pm.00001\\0.09081\pm.00006\\}{p=3\times 10^{-13}}
&\zp{1.62396\pm.00001\\0.09062\pm.00007\\}{p=0.08}
&\zp{1.62397\pm.00001\\0.09054\pm.00010\\}{p=0.41}
&\zp{1.62398\pm.00002\\0.09048\pm.00015\\}{p=0.43}
&\zp{1.62399\pm.00003\\0.0904\pm.0002\\}{p=0.56}
&\zp{1.62400\pm.00005\\0.0902\pm.0005\\}{\text{$p$ undefined}}
\end{tabular}
\end{ruledtabular}
\caption{Estimates of the 3D LERW dimension from the data.  Different sets of system size $L$ were used; the fit in the first data column was for $L\in\{2^8,2^9,2^{10},2^{11},2^{12},2^{13},2^{14}\}$, while the last column used only $L\in\{2^{13},2^{14}\}$.  The fits shown here are least-squares fits of $\log({\mathbb E}[\text{loop length}])$ to functions of the form $z \log L + a$, where we used $10^9$ data points for each system size $L$.  (The parameter $a$ depends upon the lattice, but $z$ is the same for both lattices.)  The error bars given are the 95\% confidence intervals ($\pm1.96$ standard deviations) of the fitted parameters ($z$ and~$a$).  For each such fit we did a $\chi^2$ test, and give the $p$-value of the $\chi^2$ statistic.}
\label{tbl:least-sqrs}
\end{table*}

We collected an enormous amount of data (using high performance
computing clusters), e.g., $10^9$ data
points for $L=16384$ for two different lattices,
where each data point requires $\approx L^2$ random walk steps, for a
total of $\approx 5\times 10^{17}$ random walk steps.  For the results to be
meaningful, we require a high-quality random number generator,
and we used one based on the advanced encryption standard (AES-256),
which has been found to have excellent statistical properties
\cite{HW:AES,TestU01}.

Following Agrawal and Dhar \cite{agrawal-dhar}, we used a hashtable of
points visited by the loop-erased random walk to identify newly
created loops.  The storage requirements are then order
$L^{1.6240\ldots}$ rather than order $L^3$.  However, unlike the
hashtables in other simulations \cite{agrawal-dhar,grassberger},
rather than use linked lists in the event that two different points
on the LERW accidentally hash to the same entry in the hashtable, we
used an ``open address'' hash table, since they have less data
structure overhead \cite{CLRS}.
Open address hash tables are not normally suitable when entries can be
deleted (such as when loops are erased), but in
the case of LERW simulations, we may delete the points in an erased
loop in reverse chronological order, and in this case open addressing
works with deletions.

\section{Estimate of dimension}

\begin{figure}[htp]
\psfrag{X}[rb][rt]{$\frac{\text{length}}{L^{1.6240}}$}
\psfrag{Y}[rt][rt][0.8]{probability density}
\subfigure[\ PDF of loop length (cubic lattice, cubic torus)]{
\includegraphics[width=\columnwidth]{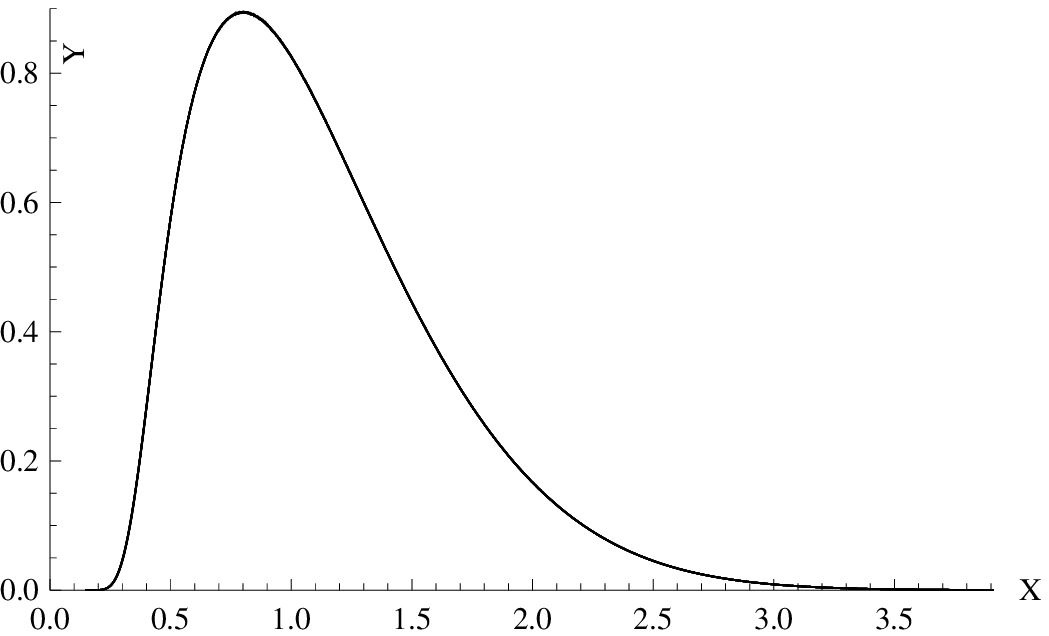}
}
\\[3.5pt]
\subfigure[\ PDF of loop length (face-centered cubic lattice, skew torus)]{
\includegraphics[width=\columnwidth]{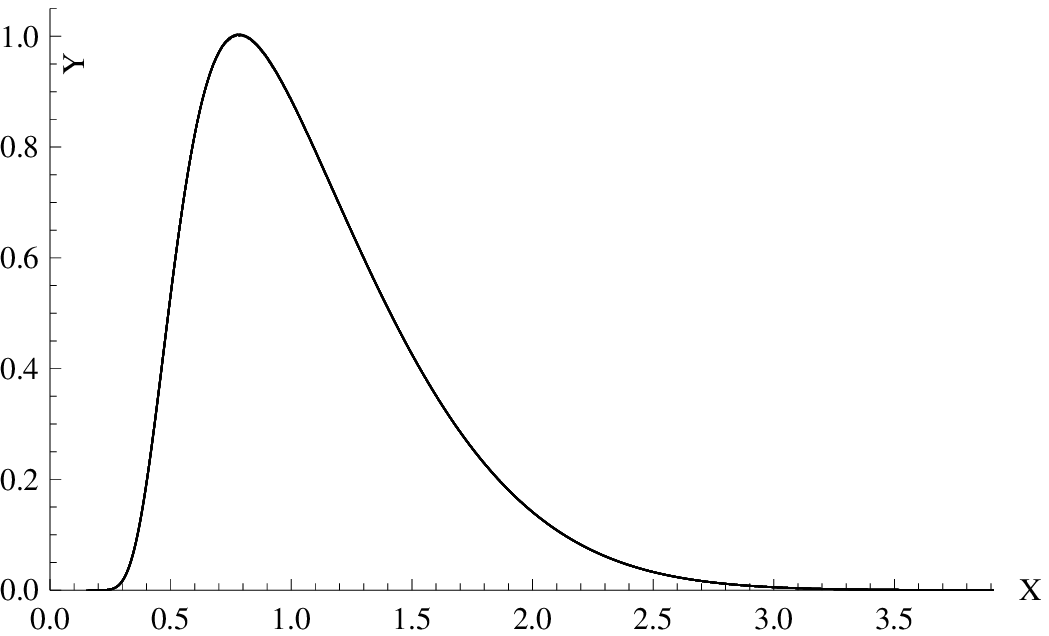}
}
\caption{Probability density function for the length of the loop-erased random loop on the $L\times L\times L$ cubic lattice (top) and face-centered cubic lattice (bottom).  For both lattices, data is shown for $L=2^9,2^{10},2^{11},2^{12},2^{13},2^{14}$.  For each $L$ we effectively show a histogram of the loop lengths; the intervals are very short, but nonetheless contain many data points, so the histograms appear to be curves.  Here the loop lengths have been scaled down by a factor of $L^{1.6240}$ to compare the data for different system sizes.  At this scale, the curves for the six different $L$'s are indistinguishable, and appear to be one curve.  The differences between the cubic and FCC curves arise because the FCC simulations were done on a skew torus.}
\label{fig:pdf}
\end{figure}

Table~\ref{tbl:least-sqrs} summarizes some simple least-squares fits
of the data to estimate the dimension, which suggest an exponent of
$1.6240$.  In Fig.~\ref{fig:pdf} we show our simulation data for the
length of the noncontractible loop in a histogram type format.  It is
evident from this figure that the exponent $1.6240$ is at least
approximately correct.

\begin{figure}[t]
\centering
\psfrag{q}[rb][ct][0.8]{$q$}
\psfrag{dimension}[ct][ct][0.8]{dimension estimate}
\psfrag{ratio}[ct][ct][0.8]{from quantile ratio}
\psfrag{16384}[cb][cb][0.6][18]{$L=16384$ vs $4096$}
\psfrag{8192}[cb][cb][0.6][43]{$L=8192$ vs $2048$}
\psfrag{4096}[cb][cb][0.6][31]{$L=4096$ vs $1024$}
\psfrag{2048}[cb][cb][0.6][40]{$L=2048$ vs $512$}
\subfigure[\ Estimate of dimension from ratios of quantiles (cubic lattice)]{
\includegraphics[width=\columnwidth]{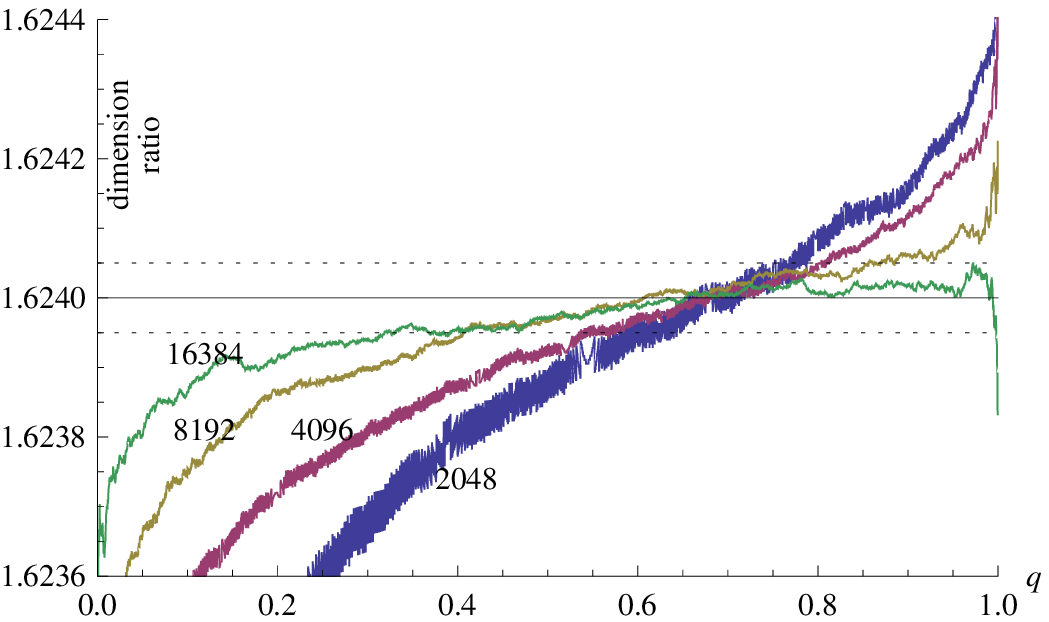}
}
\\
\subfigure[\ Estimate of dimension from ratios of quantiles (FCC lattice)]{
\psfrag{16384}[cb][cb][0.6][10]{$L=16384$ vs $4096$}
\psfrag{8192}[cb][cb][0.6][22]{$L=8192$ vs $2048$\ }
\psfrag{4096}[cb][cb][0.6][38]{$L=4096$ vs $1024$}
\psfrag{2048}[cb][cb][0.6][37]{$L=2048$ vs $512$}
\includegraphics[width=\columnwidth]{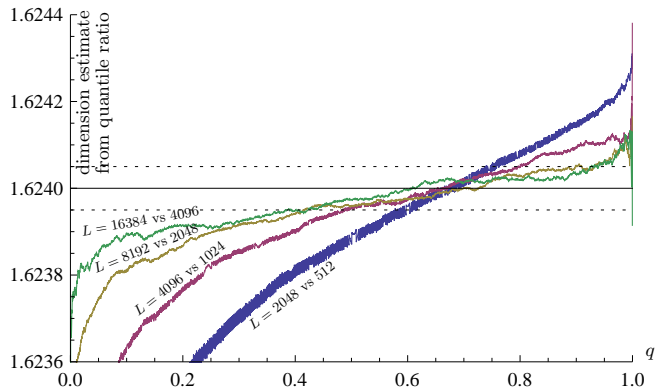}
}
\caption{(Color online)  Here we plot the quantile ratios $\log(Q_q(L)/Q_q(L/4))/\log(4)$ versus $q$ for $L=2^{11},2^{12},2^{13},2^{14}$ for the cubic lattice (top) and face-centered cubic lattice (bottom).
The saw-tooth pattern, which is evident for $L=2^{11}$ and still visible for $L=2^{12}$, arises because the loop length is integer-valued.  The saw teeth are more pronounced for the cubic lattice because it is bipartite.  The curves for larger $L$ become progressively more flat, and appear to be converging to a value in the range $1.62400\pm 0.00005$.}
\label{fig:qq}
\end{figure}

To estimate the exponent to four decimals from a plot, we need another
way to present the data.  Therefore we let $Q_q(L)$
denote the $q$\th quantile for loop length on a system of size $L$.
For example, $Q_{0.5}(16384)$ is the empirical median loop length for
the torus of order 16384.  For any $q$, we would expect $Q_q(L)$ to
take the form $a(q) L^z$ for large $L$.
More precisely, we would expect there to be a correction
term, most likely of the form $Q_q(L) = a(q) L^z +
b(q) L^y + \cdots$, where $y$ might be related to the exponent for a close
encounter of the LERW path with itself.  Then 
$$\frac{\log(Q_q(L_1)/Q_q(L_2))}{\log(L_1/L_2)} = z + \frac{b(q)/a(q)}{\log(L_1/L_2)} (L_1^{y-z}-L_2^{y-z}) + \cdots$$
In Fig.~\ref{fig:qq}, we
plot these ratios of empirical quantiles to estimate the dimension.
It appears that $b(q)$ is 
negative for small $q$ and positive for large $q$, taking the value $0$ near $q=0.7$.
For this $q$ the first correction term is $0$, so the quantile
ratios more precisely give $z$.  The quantile ratios (as a function of
$q$) appear to converge to a horizontal line at a geometric rate,
in agreement with the formula.  
Additional analysis of the data suggests that the correction exponent
$y-z$ is approximately in the range $-0.8$ to $-0.85$.
Based on these plots, together with
the least-squares fits, we judge that the dimension $z$ is likely to
be $1.62400\pm 0.00005$.

\section{Tail behavior}

Also of interest are quantities such as the probability that a
loop-erased path is unusually short.  In two dimensions it was
recently shown that the probability that a LERW path is shorter than
$\lambda$ times its expected length decays exponentially fast in
$\lambda^{-4/5+o(1)}$ (or faster) \cite{BM}.  We estimated the
corresponding tail behavior for 3D LERW, and found that it decays
exponentially fast in $\lambda^{-\alpha}$ for $\alpha$ approximately
$0.58\pm 0.02$.  Perhaps the correct exponent $\alpha$ is the
reciprocal of the dimension.

\vspace{\baselineskip}
\noindent
\textbf{Acknowledgements.}
We thank Russ Lyons for comments on an earlier version.

\bibliography{lerw3d}

\end{document}